# Functional meta lenses for compound plasmonic vortex field generation and control

E. Prinz[1,*], G. Spektor[2,3,4,*†], M. Hartelt[1], A.K. Mahro[1], M. Aeschlimann[1] and M. Orenstein[2].

[1]*Department of Physics and Research Center OPTIMAS, University of Kaiserslautern, Erwin Schroedinger Strasse 46, 67663 Kaiserslautern, Germany.*
[2]*Department of Electrical Engineering, Technion - Israel Institute of Technology, 32000 Haifa, Israel.*
[3]*Associate of the National Institute of Standards and Technology, Time and Frequency Division, Boulder, Colorado 80305, USA (curr. address)*
[4]*Department of Physics, University of Colorado, Boulder, Colorado 80309, USA (curr. address)*

[*]E. Prinz and G. Spektor contributed equally to this work as first authors.
†Corresponding author. Email: grisha.spektor@gmail.com

**Abstract:**

Surface plasmon polaritons carrying orbital angular momentum are of great fundamental and applied interest. However, common approaches for their generation are restricted to having a weak dependence on the properties of the plasmon-generating illumination, providing a limited degree of control over the amount of delivered orbital angular momentum. Here we experimentally show that by tailoring local and global geometries of vortex generators, a change in circular polarization handedness of light imposes arbitrary large switching in the delivered plasmonic angular momentum. Using time-resolved photoemission electron microscopy we demonstrate pristine control over the generation and rotation direction of high-order plasmonic vortices. We generalize our approach to create complex topological fields and exemplify it by studying and controlling a "bright vortex", exhibiting the breakdown of a high-order vortex into a mosaic of unity-order vortices while maintaining the overall angular momentum density. Our results provide tools for plasmonic manipulation and could be utilized in lab-on-a-chip devices.

**Main:**

Surface Plasmon Polaritons (SPPs) are evanescent electromagnetic waves propagating along metal-dielectric interfaces. In recent years, their ability to carry surface-confined orbital angular momentum (OAM) and form plasmonic vortices has been of wide interest [1–7]. Understanding and controlling such vortices opens the door towards a variety of applications. Examples are the unlocking of forbidden multipolar transitions in novel light-matter interactions [8–10] and plasmonic tweezers for biological and chemical purposes [11–14]. For the latter, the trapping and rotating of both dielectric [13] and metallic [11,12,14] microparticles have been demonstrated. For the manipulation of metallic particles, tweezers utilizing plasmonic OAM have been proven to be superior compared to purely optical vortices, allowing for more precise positioning of the particles due to their azimuthal scattering forces.

Plasmonic OAM can be generated by the use of plasmonic vortex generators (PVGs), an engraved structure based on Archimedean spirals (Fig.1 (A), (B)) [1,3]. These PVGs typically receive circularly polarized light as an input and generate vortices carrying OAM. This light-spin ($S = \sigma\hbar$ with spin quantum number $\sigma = \pm 1$) to plasmonic OAM conversion (topological charge $L = l\hbar$) at the structure can be described via $l = m + \sigma$ where $l$ is the orbital angular momentum quantum number and $m$ is the geometric order of the spiral [2]. The influence of the handedness of circularly polarized light on the plasmonic momentum was termed Optical Spin Hall Effect (OSHE) [15,16]. The conversion equation implies that the generated vortices gain most of their angular momentum from the structure, while the illumination only contributes a single unit of OAM. A change in the helicity of the incident light alters the order of the vortex by 2. Therefore, in the case of high order spirals, the direction of rotation of the plasmonic vortex, and thus the sign of the topological charge, is essentially determined by the geometrical order of the structure in such a way that the state of illumination has only a mild effect.

Here we develop a versatile plasmonic topology manipulator that amplifies the role of the illumination state on the generated angular momentum carrying plasmons, creating an arbitrary OAM state per circular polarization handedness of illumination. To achieve this, we combine the chiral geometry of vortex generators with the capability of optical metasurfaces to locally shape the field. Metasurfaces have proven to be a good tool for shaping and controlling SPPs [4,16–22] and we utilize a symmetry property of a unidirectional SPP launcher [17] as a basis for our approach. The combined chiral and metasurface boundary approach allows the creation of structures with an effective chirality that depends on the illumination, resulting in a plasmonic OAM governed by $l = m_\sigma + \sigma$, where $m_\sigma$ is now an arbitrary function of the illumination handedness. In essence, each orthogonal circular polarization state of the incident light results in an arbitrary plasmonic vortex that is predetermined by the structure design.

As an example that will be discussed further below, we experimentally show a hetero-chiral structure that employs polarization of the light to completely switch (flip) the OAM of a vortex from $l = 11$ to $l = -11$, providing a total difference of $22\hbar$ (Fig.1 (C, D)). In this case, the light polarization handedness determines the direction of rotation of the high order vortex. The OAM created by the structure follows the above $l = m_\sigma + \sigma$, where $m_\sigma = 10\sigma$. Namely the resulting effective geometrical order $m_\sigma$ here is determined by the incident polarization handedness.

While earlier theoretical studies also proposed meta structures designs allowing the control of plasmonic OAM depending on the handedness of the excitation light [19,21], we present and experimentally realize a more general approach where compound fields can be generated by the modular superposition of different structures. As an example for this, we assemble a "bright vortex" generator that overlays the singularity of a high order plasmonic vortex with a high-intensity peak. Upon switching the polarization of the illumination, the vortex changes its sign while the focal peak is retained. We experimentally verify our approach using time-resolved two-photon photoemission microscopy (TR-PEEM) providing 30 nm spatial and 100 as temporal step size [6]. We record the sub-femtosecond dynamics of the plasmonic fields in our assemblies, demonstrating the control over the direction of the vortex rotation.

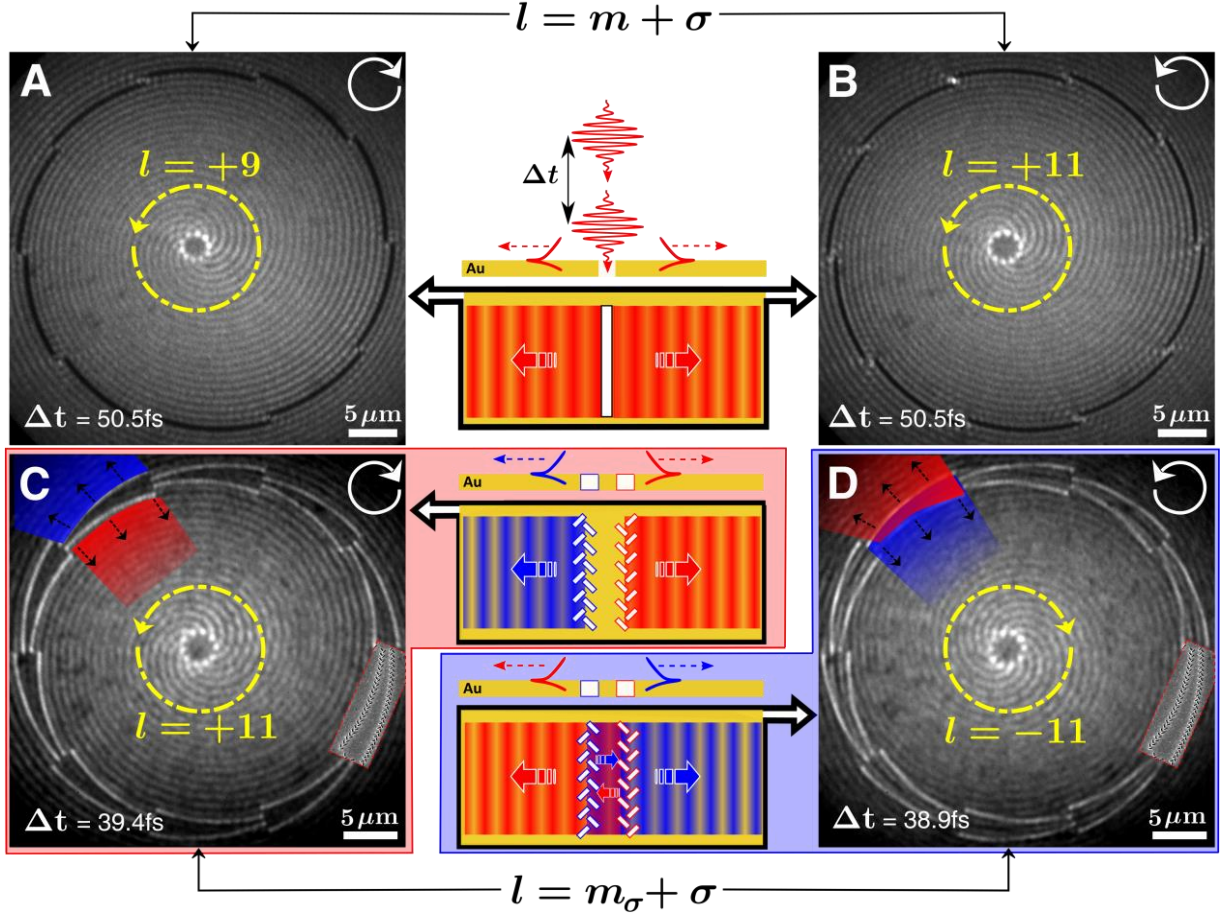

**Figure 1. TR-PEEM snapshots of simple groove and meta-slit structures.** With the slit structure, a counter-clockwise rotating vortex (indicated by the dashed yellow arrows) is created for both left (A) and right (B) handed circular polarization illumination (indicated by the white arrows in the upper right corners). Using the concept of interleaved meta-slits, the vortex flips its rotation orientation from counter-clockwise (C) to clockwise (D) for left- and right-handed circular polarization illumination respectively (See full movies in SI). The overlays and schematics are color-coded such that blue is launched from the outer meta-slits and red from the inner meta-slits. The red dashed insets in C and D are the scanning electron microscope overlays of the nanostructured metasurfaces.

A common method for the creation of higher-order plasmonic OAM is the use of grooves in the shape of segmented Archimedean spirals, which have lower azimuthal losses than continuous ones [3]. Their radius is given by $r_m(\phi) = r_i \pm \lambda_{SPP} \cdot \mathrm{mod}(m\phi, 2\pi)/2\pi$, where $\phi$ is the azimuthal angle, $\lambda_{SPP}$ is the SPP wavelength and the sign determines the handedness of the spiral. The steady state plasmonic field components inside the spiral can be approximated mathematically by Bessel functions corresponding to the plasmonic OAM [3]: $E_r \approx J'_l(k_{SPP}r) \cdot exp(il\theta)$, $E_\theta \approx J_l(k_{SPP}r) \cdot exp(il\theta)$, $E_z \approx J_l(k_{SPP}r) \cdot exp(il\theta)$. $r, \theta, z$ are the cylindrical coordinates, $J_l$ is the Bessel function of order $l$, $J_l'$ is its first derivative, $k_{SPP}$ is the plasmonic wave vector and $i$ is the imaginary unit. The approximation is valid for sufficiently large structures, with relatively low propagation losses and the reflection from the boundaries is neglected. It should be noted that, due to the subtractive nonlinear spin-orbit mixing between the circularly polarized probe-light and the plasmonic vortex field which is inherent to the TR-PEEM technique [7], the actual number of lobes in all measured vortices is given by $l_{exp} = l - \sigma_{probe}$, with $\sigma_{probe}$ the handedness of the probe-light.

To augment the capabilities of such segmented spirals using metasurfaces, we employ a configuration of double chains of orthogonal slanted nano-apertures (Fig. 1 (C, D)) [17] (see SI for the structure parameters). When illuminated with circularly polarized light, these metastructures provide unidirectional SPP excitation, with the launching direction determined by the handedness of the incident light [17,20].

The first meta-chiral augmented structure we explore is a vortex gate controlled by the circular polarization of the light. When a meta segmented spiral of the order $m = 10$ (Fig. 2) is excited by right circularly polarized light it excites only plasmons that propagate outward, away from the center of the structure, and do not form a plasmonic vortex, leaving only a weak residual field within the lens perimeter (Fig. 2 (B)). For left circularly polarized light, SPPs are launched inward, towards the center of the structure, and form a full-fledged clockwise rotating vortex of order $l = 11$ (Fig. 2 (C)).

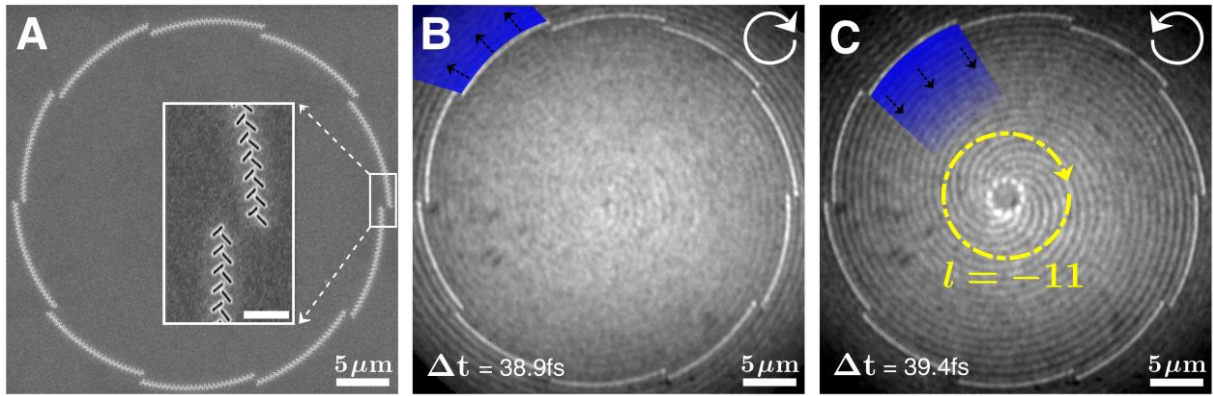

**Figure 2. Fundamental structure.** Experimental results of the meta-PVG of order 10. (A) SEM scan of the basic building block. The size of the scale bar in the inset is 1 μm. (B, C) TR-PEEM snapshots. For left circularly polarized incident light only a weak residual field penetrates the lens (B) and for the right-handed circular polarization a clockwise rotating vortex is obtained (C) The white arrows in the upper right corners indicate the handedness of the incident circularly polarized light.

To achieve a circular polarization-controlled switch, where the light polarization handedness determines the rotation direction of the plasmonic vortex, we introduce an additional meta spiral with both mirror-flipped chirality and meta-slit orientation (Fig. 1 (C, D)). The mirroring of the global chiral geometry facilitates the generation of a vortex with a topological charge of an opposite sign; whereas the mirroring of the local metasurface geometry reverses the direction of the vortex gate of this structure. Each meta-slit column is capable of unidirectional SPP excitation albeit in opposite directions (Fig. 1 (C, D)).

Illuminating this pair with one state of circular polarization selectively addresses only one excitation direction of each column. While a simple groove symmetry results in a unidirectional vortex rotation direction regardless of the excitation polarization (Fig. 1(A, B)), the incident light's handedness acts as a selector for two columns of meta-slits shaped as spirals with opposite handedness. It determines both the order and the sign of the plasmonic OAM. Dressing the meta-slit columns with segmented spiral geometries of different geometrical orders would functionalize them to launch plasmonic vortices of different topological orders to the center of the structure, depending on the incident polarization.

To demonstrate the high order switching we generated a bi-directionally rotating vortex field with a spiral pair of the geometric order 10, where the rotation direction is controlled by the light polarization (Fig.1 (C, D)). In the case of left circular polarization SPPs are launched towards the interior of the structure from the inner spiral, leading to a counter-clockwise rotation direction of the plasmonic vortex, while the SPPs created at the outer spiral propagate outwards. When illuminating the lens with right circularly polarized light, the plasmons created at the outer spiral propagate inwards and form a clockwise rotating vortex. In this case, the SPPs from the inner spiral are launched in the outside direction.

To support our experimental observations, we modeled the out-of-plane electrical plasmon field component created by a spiral pair of order 5 (Fig. 3). Since this field component is the most dominant one, it is of special interest when considering e.g. interaction with particles outside the metal layer. Full electromagnetic simulations were performed using a 3D finite difference time domain simulator by Lumerical Inc. In agreement with our experiments, the simulations reveal that the handedness of the incident polarization acts as a switch for the rotation direction of the plasmonic vortex.

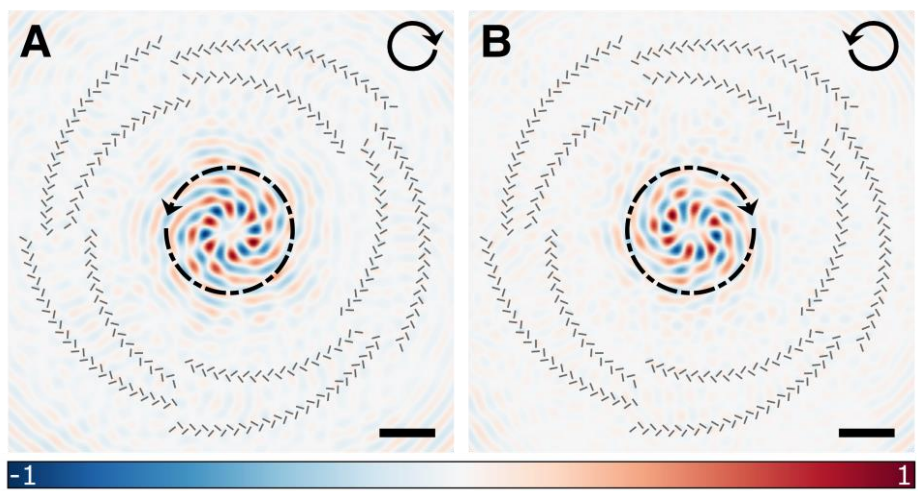

**Figure 3. Out-of-plane component FDTD simulation.** A structure of two combined meta-plasmonic vortex generators of geometric order 5 produces a vortex of order 6 with 6 red lobes (corresponding to a positive electrical field) that rotate in either counter-clockwise (A) or clockwise (B) direction (black dashed arrows), depending on the incident circular polarization (black arrows). The positions of the nano-apertures are indicated in grey. Both images depict time-snapshots of the vortex formation. The scale bars are 2 μm.

Finally, we show that our method can be generalized to assemble more complex meta-spiral structures allowing more intricate field generation and control. As an example, we constructed a "bright vortex" generator (Fig. 4) that combines the plasmonic field created by a circle (m=0, blue in Fig. 4 (A-C)) with the field that is created by a spiral of the order m=±10 (gold in Fig. 4 (A-C). As shown above, the addressed spiral depends on the handedness of the circularly polarized illumination. The resulting plasmonic field distribution recorded with TR-PEEM, considering the subtractive spin-orbit mixing [7], features a peak field amplitude located inside the phase singularity of a vortex, corresponding to

$$E_z \propto a \cdot J_0(k_{SPP}r) + b \cdot J_{10}(k_{SPP}r)\cos(\omega \cdot \Delta t \pm 10 \cdot \theta),$$

where $a$ and $b$ are electrical field amplitudes. In such a merger, strictly speaking, the singularity vanishes, and the plasmonic high order vortex field breaks down into small vortices of order unity surrounding the perimeter of the original pure vortex. Calculating the absolute value and phase distribution of the field (Fig. 4 (D, E)) shows how the combined field breaks down into this structured array of vortices (indicated by black circular arrows in Fig. 4 (D-F)). The unity order vortices correspond to minima of the absolute field strength (Fig. 4 (D)) and to singularities surrounded by a gradient of $2\pi$ in the phase distribution (Fig. 4 (E)). The out-of-plane angular momentum density of the total field on the other hand, proportional to $\vec{r} \times \vec{E} \times \vec{H}$ with respect to the origin, is merely perturbed by the additional field (Fig. 4 (F)). Thus, even though the vortex breaks in a strict mathematical sense, the angular momentum density of the compound field remains governed by the vortex contribution. The full movies in the SI show the detailed dynamics of the arrangement of the small order 1 vortices rotating in a gear-like array along the perimeter of the main lobe of the pure vortex field. Our theoretical and experimental results exhibit excellent agreement.

One possible modification of this structure design would be the removal of one of the two inner circles of the structure. In this case, the peak field amplitude at the center would only be generated for one handedness of the circularly polarized illumination and could be switched on and off easily. Similarly, by removing and adding spirals of the desired order and handedness, a great variety of plasmonic fields that can be changed on demand could be generated.

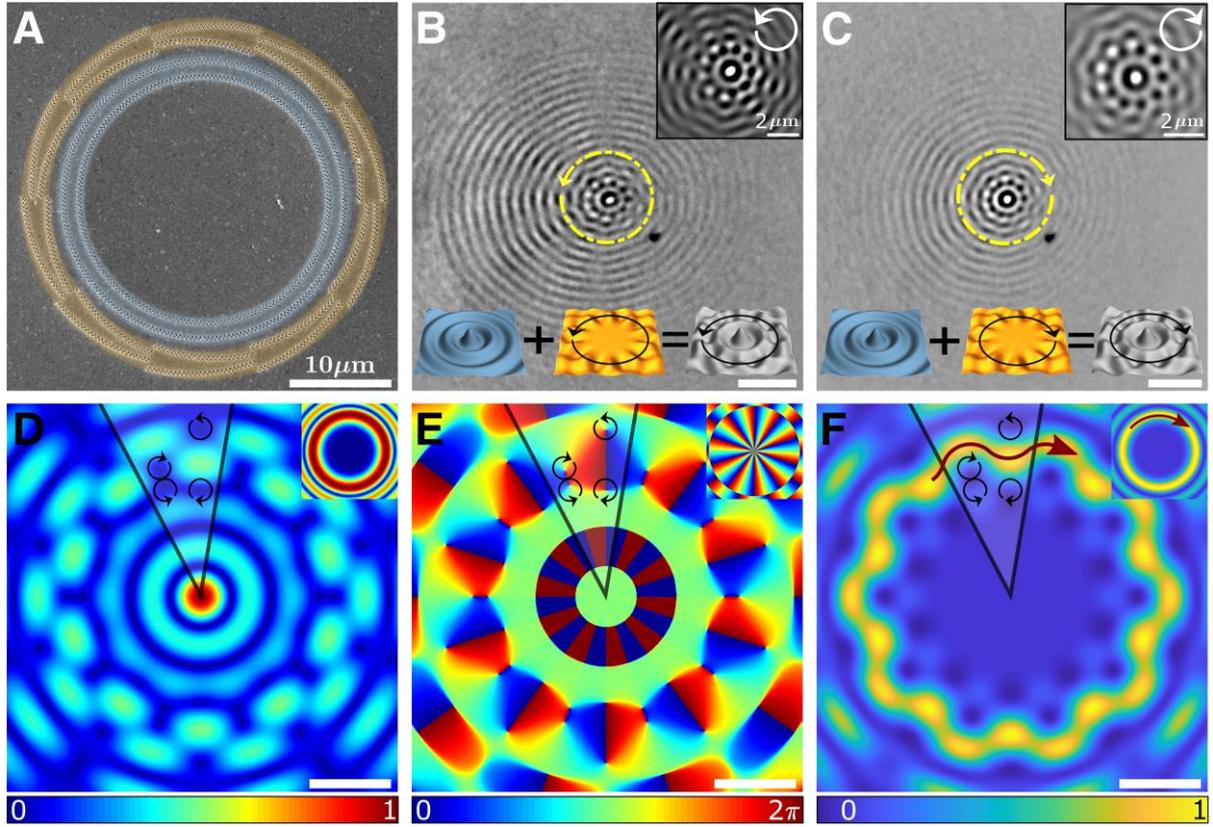

**Figure 4. "Bright vortex".** (A) SEM scan of a modular structure producing the *bright vortex* field distribution. Blue and yellow overlay regions represent the meta-spiral structures generating the bright central spot ($l = 0$, blue in (B, C)) and vortex field ($l = \pm 10 = m_\sigma$), gold in (B, C)) respectively. (B, C) Fourier-filtered experimental data snapshots of the electron emission pattern distribution that display only the signal contribution that oscillates with the frequency of the illuminating light *[7]* (thus removing the static background), the scale bar is 5 $\mu m$. Notably, a bright spot appears within the rotating field distribution. Upon changing the excitation polarization (white circular arrows), the vortex rotation (yellow arrows) is inverted while the central spot remains. The top right insets show a magnification of the bright vortex, the bottom insets show a schematic theoretical calculation of the field distribution using steady-state analytic field profiles. The resulting grey field profile matches the experimental field distribution. (D-F) Normalized theoretical modeling of a vortex of order $l = 10$ with a bright central spot ($l = 0$), absolute value (D), phase (E), and calculated angular momentum density (F) of the out-of-plane field, the scale bars are 5 $\mu m$. The combined field breaks down into a structured array of unity order vortices, indicated in (D-F) by the black circular arrows for one azimuthal unit cell. The dark red arrows in (F) indicates the contour and orientation of maximal angular momentum density. However, the overall OAM Density is still mainly governed by $l = 10$ vortex contribution (F). The insets show the intensity, phase, and angular momentum density distributions for a pure vortex of order $l = 10$.

To summarize, we have shown that combining the hetero-chirality principle with unidirectional meta-slits plasmonic launchers allows the shaping of on-demand fields with a pristine degree of control. In particular, we demonstrated that the effective chirality of the assembly changes as a function of illumination handedness, enabling the control of rotation orientation for high order vortices. The presented structures can facilitate concurrent bi-directional rotation and trapping of particles regardless of their dielectric contrast to the surrounding medium by selectively generating a peak or a dip in the plasmonic field distribution.

**Acknowledgements**

G. S. and M. O. acknowledge support from the Israeli Centers of Research Excellence "Circle of Light." We acknowledge the Micro-Nano Fabrication Unit (MNFU) Technion for support with sample fabrication, we acknowledge Dr. Guy Ankonina for the sputtering of gold. We would like to thank Or Avitan and Matan Tsipory for help with simulations. G.S. would like to acknowledge the support from the Clore Israel Foundation and the Schmidt Science Fellowship. E.P. acknowledges support from the Max Planck Graduate Center with the Johannes Gutenberg University Mainz and the University of Kaiserslautern through a PhD fellowship.

**References:**

[1] T. Ohno and S. Miyanishi, Study of surface plasmon chirality induced by Archimedes' spiral grooves, *Opt. Express* **14**, 6285–6290 (OSA, 2006).

[2] Y. Gorodetski, A. Niv, V. Kleiner, and E. Hasman, Observation of the Spin-Based Plasmonic Effect in Nanoscale Structures, *Phys. Rev. Lett.* **101**, 043903 (American Physical Society, 2008).

[3] H. Kim, J. Park, S.-W. Cho, S.-Y. Lee, M. Kang, and B. Lee, Synthesis and Dynamic Switching of Surface Plasmon Vortices with Plasmonic Vortex Lens, *Nano Letters* **10**, 529–536 (2010).

[4] C.-F. Chen, C.-T. Ku, Y.-H. Tai, P.-K. Wei, H.-N. Lin, and C.-B. Huang, Creating optical near-field orbital angular momentum in a gold metasurface, *Nano letters* **15**, 2746–2750 (ACS Publications, 2015).

[5] A. David, B. Gjonaj, and G. Bartal, Two-dimensional optical nanovortices at visible light, *Phys. Rev. B* **93**, 121302 (American Physical Society, 2016).

[6] G. Spektor, D. Kilbane, A. K. Mahro, B. Frank, S. Ristok, L. Gal, P. Kahl, D. Podbiel, S. Mathias, H. Giessen, F.-J. Meyer zu Heringdorf, M. Orenstein, and M. Aeschlimann, Revealing the subfemtosecond dynamics of orbital angular momentum in nanoplasmonic vortices, *Science* **355**, 1187–1191 (American Association for the Advancement of Science, 2017).

[7] G. Spektor, D. Kilbane, A. K. Mahro, M. Hartelt, E. Prinz, M. Aeschlimann, and M. Orenstein, Mixing the Light-spin with Plasmon-orbit by non-linear light matter interaction in gold, *Phys. Rev. X* **9**, 021031 (American Physical Society, 2019).

[8] N. Rivera, I. Kaminer, B. Zhen, J. D. Joannopoulos, and M. Soljačić, Shrinking light to allow forbidden transitions on the atomic scale, *Science* **353**, 263–269 (American Association for the Advancement of Science, 2016).

[9] C. T. Schmiegelow, J. Schulz, H. Kaufmann, T. Ruster, U. G. Poschinger, and F. Schmidt-Kaler, Transfer of optical orbital angular momentum to a bound electron, *Nature*

*Communications* **7**, 12998 (2016).

[10]  W. Cai, O. Reinhardt, I. Kaminer, and F. J. G. de Abajo, Efficient orbital angular momentum transfer between plasmons and free electrons, *Phys. Rev. B* **98**, 045424 (American Physical Society, 2018).

[11]  Z. Shen, Z. J. Hu, G. H. Yuan, C. J. Min, H. Fang, and X.-C. Yuan, Visualizing orbital angular momentum of plasmonic vortices, *Opt. Lett.* **37**, 4627–4629 (OSA, 2012).

[12]  C. Min, Z. Shen, J. Shen, Y. Zhang, H. Fang, G. Yuan, L. Du, S. Zhu, T. Lei, and X. Yuan, Focused plasmonic trapping of metallic particles, *Nature communications* **4**, 2891 (Nature Publishing Group, 2013).

[13]  W.-Y. Tsai, J.-S. Huang, and C.-B. Huang, Selective Trapping or Rotation of Isotropic Dielectric Microparticles by Optical Near Field in a Plasmonic Archimedes Spiral, *Nano Letters* **14**, 547–552 (2014).

[14]  Y. Zhang, W. Shi, Z. Shen, Z. Man, C. Min, J. Shen, S. Zhu, H. P. Urbach, and X. Yuan, A plasmonic spanner for metal particle manipulation, *Scientific reports* **5**, 15446 (Nature Publishing Group, 2015).

[15]  C. Leyder, M. Romanelli, J. P. Karr, E. Giacobino, T. C. H. Liew, M. M. Glazov, A. V. Kavokin, G. Malpuech, and A. Bramati, Observation of the optical spin Hall effect, *Nature Physics* **3**, 628–631 (2007).

[16]  N. Shitrit, I. Bretner, Y. Gorodetski, V. Kleiner, and E. Hasman, Optical Spin Hall Effects in Plasmonic Chains, *Nano Letters* **11**, 2038–2042 (2011).

[17]  J. Lin, J. P. B. Mueller, Q. Wang, G. Yuan, N. Antoniou, X.-C. Yuan, and F. Capasso, Polarization-Controlled Tunable Directional Coupling of Surface Plasmon Polaritons, *Science* **340**, 331–334 (American Association for the Advancement of Science, 2013).

[18]  S. Xiao, F. Zhong, H. Liu, S. Zhu, and J. Li, Flexible coherent control of plasmonic spin-Hall effect, *Nature Communications* **6**, 8360 (2015).

[19]  S.-Y. Lee, S.-J. Kim, H. Kwon, and B. Lee, Spin-direction control of high-order plasmonic vortex with double-ring distributed nanoslits, *IEEE Photonics Technology Letters* **27**, 705–708 (IEEE, 2015).

[20]  G. Spektor, A. David, B. Gjonaj, G. Bartal, and M. Orenstein, Metafocusing by a Metaspiral Plasmonic Lens, *Nano Letters* **15**, 5739–5743 (2015).

[21]  Q. Tan, Q. Guo, H. Liu, X. Huang, and S. Zhang, Controlling the plasmonic orbital angular momentum by combining the geometric and dynamic phases, *Nanoscale* **9**, 4944–4949 (The Royal Society of Chemistry, 2017).

[22]  Q. Tan, Z. Xu, D. H. Zhang, T. Yu, S. Zhang, and Y. Luo, Polarization-Controlled Plasmonic Structured Illumination, *Nano Letters* **20**, 2602–2608 (2020).

# Supplementary material for:

# Functional meta lenses for compound plasmonic vortex field generation and control

E. Prinz[1,*], G. Spektor[2,3,4,*], M. Hartelt[1], A.K. Mahro[1], M. Aeschlimann[1] and M. Orenstein[2].

[1]Department of Physics and Research Center OPTIMAS, University of Kaiserslautern, Erwin Schroedinger Strasse 46, 67663 Kaiserslautern, Germany.
[2]Department of Electrical Engineering, Technion - Israel Institute of Technology, 32000 Haifa, Israel.
[3]Associate of the National Institute of Standards and Technology, Time and Frequency Division, Boulder, Colorado 80305, USA (curr. address)
[4]Department of Physics, University of Colorado, Boulder, Colorado 80309, USA (curr. address)
[*]E. Prinz and G. Spektor contributed equally to this work as first authors.

**Metastructure design**

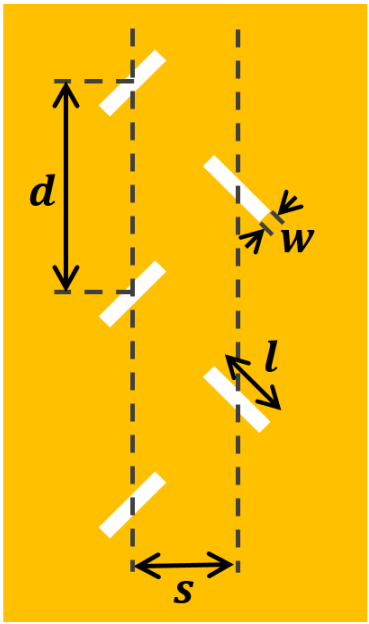

**Figure S1:** Sketch of the metastructure design (not to scale).

The following dimensions were used for all metastructures in this publication:
-Spacing between the antenna columns: s=0.196µm
-Spacing between the antennas in one column: d=0.4µm
-Width of one antenna: w=0.06µm
-Length of one antenna: l=0.3µm